\documentclass[11pt,a4paper]{article}
\usepackage[utf8]{inputenc}
\usepackage{amsmath}
\usepackage{amsfonts}
\usepackage{amssymb}
\usepackage{hyperref}
\usepackage{graphicx}
\usepackage{tikz}
\usepackage{authblk}
\usetikzlibrary{decorations.pathreplacing}
\usetikzlibrary{shapes.arrows}
\usepackage{setspace}
\usepackage[left=1.7cm,right=2cm,top=2cm,bottom=2.5cm]{geometry}
\usepackage{multicol}
\begin{document}
\title{\textbf{Spin Resolved Scattering and Spin-Orbital Entanglement in Mesoscopic Conductors}}
\author{Bibek Bhandari}
\affil{Central Department of Physics, Tribhuvan University, Kathmandu, Nepal}

\maketitle
\begin{abstract}
In this article, we have discussed about the scattering of charge and spin in spin-resolved mesoscopic systems. We have proposed a method to generate and detect the entanglement of electroninc spin and electronic orbital degrees of freedom. It is observed that in a spin-resolved mesoscopic system, all form of entanglement existing between different electronic degrees of freedom can be generated. A method to detect Bell's inequality violation in terms of zero frequency current cross-correlators has been mentioned. Further, we have also studied the effect of quantum interference between spin flip and non-spin flip scattering matrices in determining the current and noise. We have obtained the expression for shot noise of spin and charge current when an arbitrarily polarised charge current is scattered at the spin resolved potential. 
\end{abstract}

\section*{1. Introduction}
A large scale implementation of quantum computation and quantum information would require the implementation of controlled generation and detection of electronic entanglement\cite{Terhal, steane}. The entanglement of electronic degrees of freedom in mesoscopic system has attracted much interest in recent times bringing new theoretical and experimental challenges into this emerging field\cite{lebedev, Nikolai, Samuelsson, Beenakker, Burkard, Costa}. The possibility of generation and detection of entanglement between different degrees of freedom of electrons are being considered. Most of these proposals consider the entanglement between electronic spin degrees of freedom\cite{Vrijen, Recher, Patrik, Fazio} where as the others consider the entanglement between electronic orbital degrees of freedom\cite{Samuelsson, Beenakker}. However, there has been only very few works on entanglement between the spin and orbital degrees of freedom. Here we propose the generation and detection of spin-orbit entanglement in spin-resolved mesoscopic system. 

\ \\
We consider a spin-resolved mesoscopic system undergoing a spin dependent scattering determined by the non-unitarty exact scatttering matrix. The non-unitary scattering matrix preserves the spin-space entanglement and relates to the non-conserving nature of spin current. The scattering region is attached to two normal leads as shown in Fig. [1]. The spin-orbit entangled state is generated by scattering the electrons through the spin resolved potential. Some other methods like weak coupling of superconductors with the normal conductors\cite{Recher,Nikolai} has also been proposed for the generation of entangled electronic states instead of scattering at spin-resolved or spin-independent impurities. Using the scattering matrix, we can easily deduce the current and current correlations necessary for the detection of entanglement. A direct formulation of standard Bell inequality can be obtained in terms of zero frequency cross-correlations\cite{Samuelsson}. Further, a direct violation of bell inequality has been theoretically achieved in terms of shot noise produced by a tunnel barrier in a two channel conductor\cite{Beenakker}. So, we can detect the spin-orbit entanglement in a spin-resolved mesoscopic system using the current cross-correlations or the shot noise. But, before that we will be considering the generation of entangled electronic states and observe the way it is manifested in the relations for current and current correlations. We will also observe the nature and extent of spin-space entanglement in current correlations of both spin and charge current. 

\begin{tikzpicture}[scale=.5,transform shape]
\draw [ultra thick,blue] (-3,2) to[out=85,in=115] (5,2) ;
\draw[ultra thick, black] (5,2) -- (8,5);
\draw[ultra thick, black] (6,0.09) -- (10,3.8);
\draw[ultra thick, black] (6,0.09) -- (10,-3.8);
\draw[ultra thick, black] (5.5,-2.09) -- (9,-5.3);
\draw [ultra thick,blue] (5.5,-2.09) to[out=-85,in=-115] (-3,-2) ;
\draw[ultra thick, black] (-3,-2) -- (-8,-5);
\draw[ultra thick, black] (-3.8,0) -- (-9,-3);
\draw[ultra thick, black] (-3.8,0) -- (-9,3);
\draw[ultra thick, black] (-3,2) -- (-8,5);
\node at (1,0) {\textbf{\LARGE{Spin Resolved}}};
\node at (1,-1) {\textbf{\LARGE{Mesoscopic Conductor}}};
\draw [decorate,decoration={brace,amplitude=10pt},xshift=-4pt,yshift=0pt]
(-9,-6) -- (-9,6) node [black,midway,xshift=-0.6cm]{\footnotesize $~$} ;
\node at (-12,0){\textbf{\LARGE{Reservoirs}}};
\end{tikzpicture}

\ \\
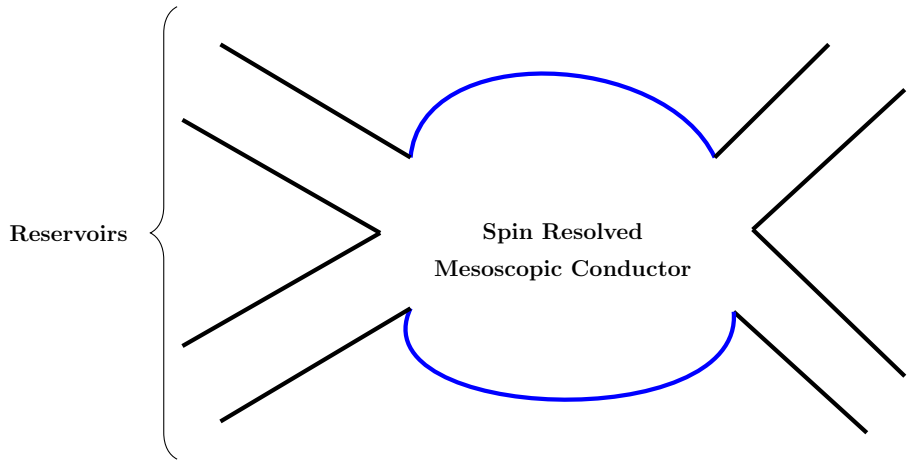
Fig. 1: A spin-dependent mesoscopic system attached to spin-resolved normal leads
\ \\
\ \\
The shot noise of charge and spin have numerous other applications in nano-sciences. The shot noise of charge current has been applied to study microscopic mechanism of charge transport and the correlation between the charges\cite{buttiker1, buttiker2,beenakker}. The spin-orbit coupled system is considered to be important ingredient in different spintronic devices\cite{zarbo,koralek}. Recent theoretical and experimental results\cite{lamacraft,belzig,mischenko,nagav,guerrero,chud} also suggests varied application of shot noise in spin dependent systems. It provides a sensitive probe to differentiate between magnetic impurities, spin flip scattering and continuous spin precession effect in transport. The spin current noise in normal conductors can be used to create fluctuating spin torque in the ferromagnetic materials. The nature of resulting magnetization noise is studied\cite{foros} by generalizing the Landauer-Buttiker formalism for spin resolved systems. The scattering matrix formulation\cite{dragomirova,foros, pareek,meair,halperin} has also been utilized to understand the effect of spin flip and non-spin flip scattering in determining spin and charge current. 
\ \\ 

The noise power spectrum of charge current in a multiterminal mesoscopic system was studied by Buttiker\cite{buttiker1,buttiker2}. We will follow Buttiker to evaluate the charge and spin transport in a spin-resolved system. It has been asserted that, in a spin resolved system the scattering matrix is physically non-unitary\cite{pareek2}. It has been seen that the charge shot noise is enhanced by the spin flip scattering on spin-orbit coupled system\cite{dragomirova}. Despite the progress in the field of shot noise, the dependence of shot noise on entanglement between different degrees of freedom hasn't been clarified. In order to obtain the necessary relations, we will express spin resolved scattering matrix in terms of space and spin dependent part of transition amplitude as done in ref. \cite{pareek2}. The scattering matrix may be written in quaternionic form as\cite{pareek};
\begin{align}
{\mathbf{S}}_{\alpha\beta}&= {\mathbf{S}}^{o}_{\alpha\beta}\otimes\mathbf{I}_{2}+{\mathbf{S}}^{x}_{\alpha\beta}\otimes{\sigma}_{x}+
{\mathbf{S}}^{y}_{\alpha\beta}\otimes{\sigma}_{y}+
{\mathbf{S}}^{z}_{\alpha\beta}\otimes{\sigma}_{z} \nonumber \\
&=\left[\begin{matrix}
\mathbf{S}^{o}_{\alpha\beta}+\mathbf{S}^{z}_{\alpha\beta} & \mathbf{S}^{x}_{\alpha\beta}-i\mathbf{S}^{y}_{\alpha\beta}  \\
\mathbf{S}^{x}_{\alpha\beta}
+i\mathbf{S}^{y}_{\alpha\beta} & \mathbf{S}^{o}_{\alpha\beta}-\mathbf{S}^{z}_{\alpha\beta}  \\
\end{matrix}\right]
\end{align}

We expressed the scattering matrix in terms of space and spin dependent part of transition amplitudes\cite{pareek2}. We may also express the scattering matrix in its block form in terms of spin flip and non-spin flip transition amplitudes.
\begin{equation}
\mathbf{S}^{\sigma\sigma '}=\left(\begin{matrix}
\mathbf{S}^{\uparrow\uparrow} & \mathbf{S}^{\uparrow\downarrow}  \\
\mathbf{S}^{\downarrow\uparrow} & \mathbf{S}^{\downarrow\downarrow}  \\
\end{matrix} \right)
\end{equation}
The later form of scattering matrix has been used to study spin resolved current and shot noise in many previous works \cite{ dragomirova, sauret,wang,belzig,nikolic}. Observing the nature of scattering matrix, the expression for shot noise can be expected to depend on initial and final spin states \cite{ dragomirova,sauret}.%

Now, the transition probability for a particular spin state is give by\cite{pareek};
\[\sum_{\alpha\beta}\mathbf{S}^{{\sigma\sigma '}^{\dagger}}_{\alpha\beta}\mathbf{S}^{\sigma\sigma '}_{\alpha\beta}=\sum_{\alpha}{R}^{\sigma \sigma '}_{\alpha\alpha}+\sum_{\alpha\neq\beta}{T}^{\sigma\sigma '}_{\alpha\beta}\]

where, $\text{R}$ is the reflection probability and $\text{T}$ is the transition probability. Interpreting equations [1] and [2], we can easily observe that the transition probabilities consists of spin-space entangled terms. The spin-space entanglement is responsible for the difference in charge current and current correlations in spin-resolved system then in spin independent one. Spin-space entanglement also explains the non-conserving nature of spin flux\cite{pareek2}.

\ \\
 We will follow the exact scattering matrix formulation to study quantum interference as well as entanglement effects in spin and charge current in a spin-resolved system. In the next section, we will deduce the expression for charge and spin current using the Landauer-Buttiker formalism\cite{buttiker1} and the exact scattering matrix\cite{pareek, pareek2}. Then, in the other two sections we will study the role of quantum entanglement between spin and space degrees of freedom and quantum interference between spin flip and non-spin flip transition amplitudes in determining self and cross correlations between currents. Finally, we will discuss about the detection of spin-space entanglement using zero frequency current cross-correlations.

\section*{2. Charge and Spin Current}
   In this section, we will use the non-unitary scattering matrix mentioned in previous section to express the charge and spin current in Landauer-Buttiker formalism. We will be using both forms of scattering matrix to explicate the nature of quantum entanglement and quantum interference in the scattered system. We take charge current as expressed in reference \cite{buttiker1} and develop thereafter using the exact non-unitary scattering matrix. We can express scattering matrix in terms of spin dependent and spin independent transition amplitudes as in equation [1]. Hence, keeping spin and space dependent part of transition amplitude, we obtain following expression for charge current;
\begin{equation}
\langle\hat{I}^{q}_{\alpha}\rangle=\frac{e}{h}\int dE \sum_{\beta\sigma, i}\Big[\left[\mathbf{N}_{\alpha}-2\text{Tr}\left[|\mathbf{S}^{0}_{\alpha\alpha}|^{2}+|\mathbf{S}^{i}_{\alpha\alpha}|^{2}\right]\right]{f}^{\sigma}_{\alpha}-2\text{Tr}\left[|\mathbf{S}^{0}_{\alpha\beta}|^{2}+|\mathbf{S}^{i}_{\alpha\beta}|^{2}\right]{f}^{\sigma}_{\beta}\Big]
\end{equation}
The spin resolved charge current involves neither the quantum interference between spin flip and non-spin flip transition amplitudes nor spin-space entanglement. Since the entangled terms occur only in the off-diagonal terms of transition probability, the average value of charge current does not involve entanglement effects. Even though the scattering matrix is taken non-unitary, the general form of average charge current remains the same, maintaining the conservation of total charge. The expressions for average spin current in transverse and longitudinal direction has been obtained in ref. \cite{pareek} and the spin current along the spin quantisation and transverse direction with the role of quantum interference has been studied. It has been observed that the spin current cannot be separated into equilibrium and non-equilibrium parts, even in a time reversal symmetric system.
The spin current along longitudinal direction is observed to be devoid of quantum interference between spin flip and non-spin flip transition amplitudes.

Now, the average spin current along longitudinal direction can be expressed as;
\begin{align}
\langle{I}^{s, z}_{\alpha}\rangle
&=\frac{1}{4\pi}\int dE \sum_{\beta \sigma} 2\Big[\text{Tr}\left[-\mathbf{S}^{{0}^{\dagger}}_{\alpha
\alpha}\mathbf{S}^{z}_{\alpha\alpha}-\mathbf{S}^{{z}^
{\dagger}}_{\alpha\alpha}\mathbf{S}^{0}_{\alpha\alpha}-
i\mathbf{S}^{{y}^{\dagger}}_{\alpha\alpha}\mathbf{S}^{x}
_{\alpha\alpha}+i\mathbf{S}^{{x}^{\dagger}}_{\alpha
\alpha}\mathbf{S}^{y}_{\alpha\alpha}
\right]{f}^{\sigma}_{\alpha} \nonumber \\ &+\text{Tr}\left[-\mathbf{S}^{{0}^{\dagger}}_{\alpha
\beta}\mathbf{S}^{z}_{\alpha\beta}-\mathbf{S}^{{z}^
{\dagger}}_{\alpha\beta}\mathbf{S}^{0}_{\alpha\beta}-
i\mathbf{S}^{{y}^{\dagger}}_{\alpha\beta}\mathbf{S}^{x}
_{\alpha\beta}+i\mathbf{S}^{{x}^{\dagger}}_{\alpha
\beta}\mathbf{S}^{y}_{\alpha\beta}
\right]{f}^{\sigma}_{\beta}\Big]
\end{align}
The average spin current in longitudinal direction does not involve quantum interference between spin flip and non spin-flip scattering amplitudes \cite{pareek} though the spin current is spin-space entangled. As observed from equation [8], we cannot separate the spin-dependent and space-dependent degrees of freedom. The expression for spin current suggests the existence of entanglement even if there is no interference between spin-flip and non-spin flip transition amplitudes. Also, even in equilibrium condition we can observe that the spin current does not vanish since we are not able to write spin current in terms of difference of distribution function.\\
\ \\
The average spin current along transverse X- direction is given by;
\begin{equation}
\langle{I}^{s, x}_{\alpha}\rangle=\frac{1}{2\pi}\int dE \sum_{\beta\sigma}\left[\text{Tr}\left[-\mathbf{S}^{{0}^{\dagger}}_{\alpha
\beta}\mathbf{S}^{x}_{\alpha\beta}-\mathbf{S}^{{x}^
{\dagger}}_{\alpha\beta}\mathbf{S}^{0}_{\alpha\beta}-
i\mathbf{S}^{{z}^{\dagger}}_{\alpha\beta}\mathbf{S}^{y}
_{\alpha\beta}+i\mathbf{S}^{{y}^{\dagger}}_{\alpha
\beta}\mathbf{S}^{z}_{\alpha\beta}\right]{f}^{\sigma}_{\beta}\right]
\end{equation}
And the average spin current along transverse Y- direction can be expressed as;
\begin{equation}
\langle{I}^{s, y}_{\alpha}\rangle=\frac{1}{2\pi}\int dE \sum_{\beta\sigma}\left[\text{Tr}\left[-\mathbf{S}^{{0}^{\dagger}}_{\alpha\beta}\mathbf{S}^{y}
_{\alpha\beta}+i\mathbf{S}^{{z}^{\dagger}}_{\alpha
\beta}\mathbf{S}^{x}_{\alpha\beta}-i\mathbf{S}^{{x}^
{\dagger}}_{\alpha\beta}\mathbf{S}^{z}_{\alpha\beta}-\mathbf{S}^{{y}^{\dagger}}_{\alpha
\beta}\mathbf{S}^{0}_{\alpha\beta}\right]{f}^{\sigma}_{\beta}\right]
\end{equation}
The expression for average spin current along transverse X and Y direction consists of quantum interference terms in between spin-flip and non spin-flip transition amplitudes. Also, entanglement in terms of spin dependent and independent terms of transition amplitudes can be observed in [9] and [10]. We find that spin space entanglement along transverse X-direction consists of X- component of spin dependent part of transition amplitude and the spin space entanglement along Y-direction consists of Y-component of spin dependent part of transition amplitude. So, the average spin current along transverse direction keeps the entanglement between space and corresponding transverse component of spin part. The other interfering term is in between the z-component and the remaining component of spin dependent transition amplitude. The entanglement between different electronic degrees of freedom observed in the case of charge and spin current shall survive even for a mesoscopic system determined by unitary scattering matrix.

\section*{3. Spin and Charge Noise in Terms of Spin Dependent and Space Dependent Transition Amplitudes}
In this section, we will determine the shot noise of spin and charge current when a beam of electrons are scattered by the spin resolved scattering potential. We know that the shot noise for charge current and spin-resolved charge current for unpolarized as well as polarized charge transport has been equally worked out. We will follow the Buttiker's approach\cite{buttiker1} in determining the spin dependent noise in terms of spin-resolved transition amplitudes. Firstly, let's write the expressions for shot noise in terms of spin and charge current. In doing so we will use equations [3] and [7] for charge and spin current and follow Buttiker\cite{buttiker1} to write the correct representation for charge and spin shot noise.The spin resolved noise, $\boldsymbol{\mathcal{S}}^{\sigma\sigma '}_{\alpha\beta}(t-{t}')=\frac{1}{2}\langle\delta{I}^{\sigma}_{\alpha}(t)\delta{I}^{\sigma '}_{\beta}(t')+\delta{I}^{\sigma '}_{\beta}(t')\delta{I}^{\sigma}_{\alpha}(t)\rangle$, where $\delta{I}^{\sigma}_{\alpha}(t)={I}^{\sigma}_{\alpha}(t)-\langle{I}^{\sigma}_{\alpha}(t)\rangle$, can be easily deduced. The expression for charge and spin current as expressed in equations [3] and [7] can be utilized to represent noise power spectrum for charge and spin as;
\begin{equation}
\boldsymbol{\mathcal{S}}^{\sigma\sigma ', s}_{\alpha\beta}=\frac{h}{{16\pi}^{2}}\int dE \sum_{\gamma\delta\sigma\sigma '}\left[{f}^{\sigma}_{\gamma}(1-{f}^{\sigma '}_{\delta})+{f}^{\sigma '}_{\delta}(1-{f}^{\sigma}_{\gamma})\right]\text{Tr}\left[ {\mathbf{\Gamma}^{i}_{\gamma\delta}}(\alpha,E){{\mathbf{\Gamma}}^{i}_{\delta\gamma}}(\beta,E)\right]
\end{equation}
\begin{equation}
\boldsymbol{\mathcal{S}}^{\sigma\sigma ', q}_{\alpha\beta}=\frac{{e}^{2}}{{h}}\int dE \sum_{\gamma\delta\sigma \sigma '} \left[{f}^{\sigma}_{\gamma}(1-{f}^{\sigma '}_{\delta})+{f}^{\sigma '}_{\delta}(1-{f}^{\sigma}_{\gamma})\right] \text{Tr}\left[\mathbf{M}_{\gamma\delta}(\alpha,E))\mathbf{M}_{\delta\gamma}(\beta,E)\right]
\end{equation}

The relations providing the expressions for $\boldsymbol{M}$ and $\boldsymbol{\Gamma}$ are given in Apppenix A.  Similar expressions for shot noise has been previously mentioned by Meair, Stano and Jacquod\cite{meair} in calculating the spin decohering transport and by Foros et. al. in calculating magnetization noise\cite{foros}. It should be noted that, since the transition probability for different spin transitions are different, hence the spin current will vary between spin states. The above proposition also alters the spin current observed in different contacts. Firstly we will calculate the shot noise associated with charge current and then proceed to spin current. We evaluate the trace of $\left[\mathbf{M}_{\gamma\delta}(\alpha,E))\mathbf{M}_{\delta\gamma}(\beta,E)\right]$, needed to determine the self and cross correlations in charge current. Under equilibrium condition, the chemical potential of different leads are the same. Recently, it has been mentioned that indistinguishability is significant in determining the entanglement and the transfer of entanglement\cite{omar}. To incorporate both the phenomena on a single footing, we have expressed entanglement in terms of scattering amplitudes and the indistinguishability has been incorporated in the fermi distribution functions. The correlation could provide an important resource in quantum information processing employing spin characteristics.  

 Firstly, let us calculate the equilibrium part of shot noise of charge  current. Under equilibrium condition $(\alpha=\beta)$, we get following expression for equilibrium shot noise;
\begin{align}
\mathbf{\mathcal{S}}^{\sigma\sigma ', q}_{\alpha\alpha}
&=\frac{2{e}^{2}}{h} \int dE\Bigg[ \sum_{i, \sigma \sigma '}\bigg[\frac{\mathbf{N}_{\alpha}}{2}- 2\text{Tr}\Big[|\mathbf{S}^{0}_{\alpha\alpha}|^{2}+|\mathbf{S}^{i}_{\alpha\alpha}|^{2}\Big]\bigg]\left[{f}^{\sigma}_{\alpha}(1-{f}^{\sigma '}_{\alpha})+{f}^{\sigma '}_{\alpha}(1-{f}^{\sigma}_{\alpha})\right]\nonumber \\
&+\sum_{\gamma\delta\sigma\sigma '}\left[{f}^{\sigma}_{\gamma}(1-{f}^{\sigma '}_{\delta})+{f}^{\sigma '}_{\delta}(1-{f}^{\sigma}_{\gamma})\right]\Big[\text{Tr}\Big[\mathbf{Q}_{\alpha\gamma\delta, \alpha\delta\gamma}+\mathbf{R}_{\alpha\gamma\delta, \alpha\delta\gamma}\Big]\Big]\Bigg]
\end{align}

where,
\begin{align}
\mathbf{Q}_{\alpha\gamma\delta, \beta\delta\gamma}=
&\sum_{i,j=\{\{0,z\},\{x,y\}\}}\mathbf{S}^{{i}^{\dagger}}_{\alpha\gamma}\mathbf{S}^{i}_{\alpha\delta}\mathbf{S}^{{j}^{\dagger}}_
{\beta\delta}\mathbf{S}^{j}_{\beta\gamma}+\sum_{i,j=\{0,x,y, z\};i\neq j}\mathbf{S}^{{i}^{\dagger}}_{\alpha\gamma}\mathbf{S}^{j}_{\alpha\delta}\mathbf{S}^{{j}^{\dagger}}_
{\beta\delta}\mathbf{S}^{i}_{\beta\gamma}+\sum_{i,j=\{0,z\};i\neq j}\mathbf{S}^{{i}^{\dagger}}_{\alpha\gamma}\mathbf{S}^{j}_{\alpha
\delta}\mathbf{S}^{{i}^{\dagger}}_
{\beta\delta}\mathbf{S}^{j}_{\beta\gamma}\nonumber \\
&-\sum_{i,j=\{x,y\};i\neq j}\mathbf{S}^{{i}^{\dagger}}_{\alpha\gamma}\mathbf{S}^{j}_{\alpha
\delta}\mathbf{S}^{{i}^{\dagger}}_
{\beta\delta}\mathbf{S}^{j}_{\beta\gamma}+i\sum\limits_{\substack{i,j,k,l=\{0,x,y,z\};{{i,l}=\{\{0,z\},\{x,y\}\}}\\{i\neq j\neq k\neq l}}}\mathbf{S}^{{i}^{\dagger}}_{\alpha\gamma}\mathbf{S}^{j}_{\alpha
\delta}\mathbf{S}^{{k}^{\dagger}}_
{\beta\delta}\mathbf{S}^{l}_{\beta\gamma}\epsilon_{xyz}
\end{align}

and

\begin{align}
\mathbf{R}_{\alpha\gamma\delta, \beta\delta\gamma}=
&\sum_{i,j=\{\{0,x\},\{0,y\},\{x,z\},\{y,z\}\};i\neq j}\mathbf{S}^{{i}^{\dagger}}_{\alpha\gamma}\mathbf{S}^{i}_{\alpha
\delta}\mathbf{S}^{{j}^{\dagger}}_
{\beta\delta}\mathbf{S}^{j}_{\beta\gamma}+\sum_{i,j=\{\{0,x\},\{0,y\}\};i\neq j}\mathbf{S}^{{i}^{\dagger}}_{\alpha\gamma}\mathbf{S}^{j}_{\alpha
\delta}\mathbf{S}^{{i}^{\dagger}}_
{\beta\delta}\mathbf{S}^{j}_{\beta\gamma}\nonumber \\
&-\sum_{i,j=\{\{x,z\},\{y,z\}\};i\neq j}\mathbf{S}^{{i}^{\dagger}}_{\alpha\gamma}\mathbf{S}^{j}_{\alpha
\delta}\mathbf{S}^{{i}^{\dagger}}_
{\beta\delta}\mathbf{S}^{j}_{\beta\gamma}+i\sum\limits_{\substack{{{i,j,k,l=\{0,x,y,z\}\text{either}~{j}=\{x,y\}~or~ k=\{x,y\}}}\\{{i\neq j\neq k\neq l}}}}\mathbf{S}^{{i}^{\dagger}}_{\alpha\gamma}\mathbf{S}^{j}_{\alpha
\delta}\mathbf{S}^{{k}^{\dagger}}_
{\beta\delta}\mathbf{S}^{l}_{\beta\gamma}\epsilon_{xyz}
\end{align}

To avoid confusion, we want to mention that the notation $i,j=\{0,z\}$ used in equations [10] and [11] implies $i=0~\text{or}~z$, and $j=0~\text{or}~z$. The expression for noise spectra for charge current suggests that the spin-space entanglement effects are present through the transport like fluctuations. Hence, unlike the charge current, the current noise is influenced by entanglement between different electronic degrees of freedom. If the scattering matrix were taken to be unitary, the equilibrium fluctuations would have been devoid of entanglement and interference effects though the transport type fluctuations would still carry entangled electrons. The charge current and corresponding noise are conserved straightforwardly in the absence of interference and entanglement effects. Moreover, we can see different patterns of interferece effect in equations [15] and [16]. The first term in the expression for $\mathbf{Q}$ and $\mathbf{R}$ produces no interference between any of the electronic degrees of freedom, though the term clearly exhibits interference between lead terminals. Further, all other terms are observed to exhibit interference between the spin-space or the spin-spin degrees of freedom. As the interference results in a non-separability of spin and space degrees of freedom, it suggests that the spin-space entanglement is the primary ingredient in the expression for shot noise in a spin resolved mesoscopic system. Now, the cross correlation of charge current can be expressed as;
\begin{align}
\mathbf{\mathcal{S}}^{\sigma\sigma ', q}_{\alpha\beta, \alpha\neq\beta}
&=\frac{2{e}^{2}}{h} \int dE \Bigg[\sum_{i, \sigma \sigma '}\Bigg[\bigg[\text{Tr}\Big[|\mathbf{S}^{0}_{\beta\alpha}|^{2}+|\mathbf{S}^{i}_{\beta\alpha}|^{2}\Big]\bigg]\left[{f}^{\sigma}_{\alpha}(1-{f}^{\sigma '}_{\alpha})+{f}^{\sigma '}_{\alpha}(1-{f}^{\sigma}_{\alpha})\right]-\text{Tr}\Big[|\mathbf{S}^{0}_{\alpha\beta}|^{2}+|\mathbf{S}^{i}_{\alpha\beta}|^{2}\Big]\nonumber \\
&\left[{f}^{\sigma}_{\beta}(1-{f}^{\sigma '}_{\beta})+{f}^{\sigma '}_{\beta}(1-{f}^{\sigma}_{\beta})\right]\Bigg]
 +\sum_{\gamma\delta\sigma\sigma '}\left[{f}^{\sigma}_{\gamma}(1-{f}^{\sigma '}_{\delta})+{f}^{\sigma '}_{\delta}(1-{f}^{\sigma}_{\gamma})\right]\Big[\text{Tr}\Big[\mathbf{Q}_{\alpha\gamma\delta, \beta\delta\gamma}+\mathbf{R}_{\alpha\gamma\delta, \beta\delta\gamma}\Big]\Big]\Bigg]
\end{align}
Both the equilibrium and non-equilibrium parts of distribution function are presented in equation [12]. This suggests that the separation of noise spectra into self and cross correlation does not necessarily separate it into equilibrium and non-equilibrium terms. Now, we express the self-correlation between spin currents along longitudinal direction as;
\begin{align}
\mathbf{\mathcal{S}}^{\sigma\sigma ', z}_{\alpha\alpha}
&=\frac{h}{8{\pi}^{2}}\int dE \Bigg[\sum_{\sigma\sigma '}\left[{f}^{\sigma}_{\alpha}(1-{f}^{\sigma '}_{\alpha})+{f}^{\sigma '}_{\alpha}(1-{f}^{\sigma}_{\alpha})\right]\bigg[\frac{\mathbf{N}_{\alpha}}{2}+2\text{Tr}\Big[|\mathbf{S}^{x}_{\alpha\alpha}|^{2}+|\mathbf{S}^{y}_{\alpha\alpha}|^{2}-|\mathbf{S}^{0}_{\alpha\alpha}|^{2}-|\mathbf{S}^{z}_{\alpha\alpha}|^{2}\Big]\bigg]\nonumber \\
&+\sum_{\gamma\delta\sigma\sigma '}\left[{f}^{\sigma}_{\gamma}(1-{f}^{\sigma '}_{\delta})+{f}^{\sigma '}_{\delta}(1-{f}^{\sigma}_{\gamma})\right]\Big[\text{Tr}\Big[\mathbf{Q}_{\alpha\gamma\delta, \alpha\delta\gamma}-\mathbf{R}_{\alpha\gamma\delta, \alpha\delta\gamma}\Big]\Big]\Bigg]
\end{align}
Moreover, the cross-correlation can be obtained as;
\begin{align}
\mathbf{\mathcal{S}}^{\sigma\sigma ', z}_{\alpha\beta, \alpha\neq\beta}
&=\frac{h}{8{\pi}^{2}}\int dE \Bigg[\sum_{\sigma\sigma '}\left[{f}^{\sigma}_{\alpha}(1-{f}^{\sigma '}_{\alpha})+{f}^{\sigma '}_{\alpha}(1-{f}^{\sigma}_{\alpha})\right]\bigg[\text{Tr}\left[|\mathbf{S}^{x}_{\beta\alpha}|^{2}+|\mathbf{S}^{y}_{\beta\alpha}|^{2}-|\mathbf{S}^{0}_{\beta\alpha}|^{2}-|\mathbf{S}^{z}_{\beta\alpha}|^{2}\right]\bigg]\nonumber \\
&+\sum_{\sigma\sigma '}\left[{f}^{\sigma}_{\beta}(1-{f}^{\sigma '}_{\beta})+{f}^{\sigma '}_{\beta}(1-{f}^{\sigma}_{\beta})\right]\Big[\text{Tr}\left[|\mathbf{S}^{x}_{\alpha\beta}|^{2}+|\mathbf{S}^{y}_{\alpha\beta}|^{2}-|\mathbf{S}^{0}_{\alpha\beta}|^{2}-|\mathbf{S}^{z}_{\alpha\beta}|^{2}\right]\Big]\nonumber \\
&+\sum_{\gamma\delta\sigma\sigma '}\left[{f}^{\sigma}_{\gamma}(1-{f}^{\sigma '}_{\delta})+{f}^{\sigma '}_{\delta}(1-{f}^{\sigma}_{\gamma})\right]\Big[\text{Tr}\Big[\mathbf{Q}_{\alpha\gamma\delta, \beta\delta\gamma}-\mathbf{R}_{\alpha\gamma\delta, \beta\delta\gamma}\Big]\Big]\Bigg]\Bigg]
\end{align}
We observe that the spin current correlation along longitudinal direction consists of spin-space entangled as well as unentangled terms. The quantum interference can be observed in between spin and space degrees of freedom as well as between different spin degrees of freedom. We also observe that the spin correlation along longitudinal direction is related to charge correlation. Because of the separation of spin along spin-quantization axis, taken here to be Z-axis, the spin correlation along the longitudinal direction and charge current correlation differ only by in-between sign. Now, we proceed to calculate spin current self and cross correlation along transverse -X and -Y direction. 
\begin{align}
\mathbf{\mathcal{S}}^{\sigma\sigma ', x,y}_{\alpha\alpha}
&=\frac{h}{8{\pi}^{2}} \int dE\Bigg[\sum_{\sigma\sigma '}
\left[{f}^{\sigma}_{\alpha}(1-{f}^{\sigma '}_{\alpha})+{f}^{\sigma'}_{\alpha}(1-{f}^{\sigma}_{\alpha})\right]\bigg[\frac{\mathbf{N}_{\alpha}}{2}-2\text{Tr}\Big[|\mathbf{S}^{x}_{\alpha\alpha}|^{2}-|\mathbf{S}^{y}_{\alpha\alpha}|^{2}\pm|\mathbf{S}^{0}_{\alpha\alpha}|^{2}\mp|\mathbf{S}^{z}_{\alpha\alpha}|^{2}\Big]\bigg]\nonumber \\
&+\sum_{\gamma\delta\sigma\sigma '}\left[{f}^{\sigma}_{\gamma}(1-{f}^{\sigma '}_{\delta})+{f}^{\sigma '}_{\delta}(1-{f}^{\sigma}_{\gamma})\right]\Big[\text{Tr}\Big[\mathbf{S}_{\alpha\gamma\delta, \alpha\delta\gamma}\pm\mathbf{T}_{\alpha\gamma\delta, \alpha\delta\gamma}\Big]\Big]\Bigg]
\end{align}

where,
\begin{align}
\mathbf{S}_{\alpha\gamma\delta, \beta\delta\gamma}=
&\sum_{i,j=\{\{0,x\},\{x,z\}\};i\neq j}\mathbf{S}^{{i}^{\dagger}}_{\alpha\gamma}\mathbf{S}^{j}_{\alpha
\delta}\mathbf{S}^{{i}^{\dagger}}_
{\beta\delta}\mathbf{S}^{j}_{\beta\gamma}-\sum_{i,j=\{\{0,y\},\{y,z\}\};i\neq j}\mathbf{S}^{{i}^{\dagger}}_{\alpha\gamma}\mathbf{S}^{j}_{\alpha
\delta}\mathbf{S}^{{i}^{\dagger}}_
{\beta\delta}\mathbf{S}^{j}_{\beta\gamma}\nonumber \\
&+\sum_{i,j=\{\{0,x\},\{y,z\}\};i\neq j}\mathbf{S}^{{i}^{\dagger}}_{\alpha\gamma}\mathbf{S}^{i}_{\alpha
\delta}\mathbf{S}^{{j}^{\dagger}}_
{\beta\delta}\mathbf{S}^{j}_{\beta\gamma}-\sum_{i,j=\{\{0,y\},\{x,z\}\};i\neq j}\mathbf{S}^{{i}^{\dagger}}_{\alpha\gamma}\mathbf{S}^{i}_{\alpha
\delta}\mathbf{S}^{{j}^{\dagger}}_
{\beta\delta}\mathbf{S}^{j}_{\beta\gamma}+i\nonumber \\
&\Bigg[\sum\limits_{\substack{{{i,j,k,l=\{0,x,y,z\};{{i,l}=\{\{x,0\},\{y,z\}\}}}}\\{{i\neq j\neq k\neq l}}}}\mathbf{S}^{{i}^{\dagger}}_{\alpha\gamma}\mathbf{S}^{j}_{\alpha
\delta}\mathbf{S}^{{k}^{\dagger}}_
{\beta\delta}\mathbf{S}^{l}_{\beta\gamma}-\sum\limits_{\substack{{{i,j,k,l=\{0,x,y,z\};{{i,l}=\{\{x,z\},\{y,0\}\}}}}\\{{i\neq j\neq k\neq l}}}}\mathbf{S}^{{i}^{\dagger}}_{\alpha\gamma}\mathbf{S}^{j}_{\alpha
\delta}\mathbf{S}^{{k}^{\dagger}}_
{\beta\delta}\mathbf{S}^{l}_{\beta\gamma}\Bigg]\epsilon_{xyz}
\end{align}

and
\begin{align}
\mathbf{T}_{\alpha\gamma\delta, \beta\delta\gamma}=
&\sum_{i,j=\{0,x,y,z\}}\mathbf{S}^{{i}^{\dagger}}_{\alpha\gamma}\mathbf{S}^{j}_{\alpha
\delta}\mathbf{S}^{{j}^{\dagger}}_
{\beta\delta}\mathbf{S}^{i}_{\beta\gamma}-\sum_{i,j=\{\{0,z\},\{x,y\}\};i\neq j}\mathbf{S}^{{i}^{\dagger}}_{\alpha\gamma}\mathbf{S}^{i}_{\alpha
\delta}\mathbf{S}^{{j}^{\dagger}}_
{\beta\delta}\mathbf{S}^{j}_{\beta\gamma}-\sum_{i,j=\{0,z\};i\neq j}\mathbf{S}^{{i}^{\dagger}}_{\alpha\gamma}\mathbf{S}^{j}_{\alpha
\delta}\mathbf{S}^{{i}^{\dagger}}_
{\beta\delta}\mathbf{S}^{j}_{\beta\gamma}\nonumber \\
&+\sum_{i,j=\{x,y\};i\neq j}\mathbf{S}^{{i}^{\dagger}}_{\alpha\gamma}\mathbf{S}^{j}_{\alpha
\delta}\mathbf{S}^{{i}^{\dagger}}_
{\beta\delta}\mathbf{S}^{j}_{\beta\gamma}+i\sum\limits_{\substack{{{i,j,k,l=\{0,x,y,z\};{{i,l}=\{\{x,y\},\{z,0\}\}}}}\\{{i\neq j\neq k\neq l}}}}\mathbf{S}^{{i}^{\dagger}}_{\alpha\gamma}\mathbf{S}^{j}_{\alpha
\delta}\mathbf{S}^{{k}^{\dagger}}_
{\beta\delta}\mathbf{S}^{l}_{\beta\gamma}\epsilon_{xyz}
\end{align}
We observe all sorts of quantum interference between space and spin degrees of freedom in the expression for charge and spin noise from equations [16] and [17]. The equilibrium-like terms are free from entanglements whereas the transport like terms consists of interference between spin and space degrees of freedom and also between different leads. These interfering terms can be said responsible for mixing type conductance observed in spin-resolved system\cite{brataas}. Now, the cross-correlation between spin currents along transverse direction is given by;
\begin{align}
\mathbf{\mathcal{S}}^{\sigma\sigma ', x, y}_{\alpha\beta}
&=\frac{h}{8{\pi}^{2}} \int dE\Bigg[\sum_{\sigma\sigma '}
\left[{f}^{\sigma}_{\alpha}(1-{f}^{\sigma '}_{\alpha})+{f}^{\sigma'}_{\alpha}(1-{f}^{\sigma}_{\alpha})\right]\bigg[\Big[|\mathbf{S}^{x}_{\beta\alpha}|^{2}-|\mathbf{S}^{y}_{\beta\alpha}|^{2}\pm|\mathbf{S}^{0}_{\beta\alpha}|^{2}\mp|\mathbf{S}^{z}_{\beta\alpha}|^{2}\Big]\bigg]\nonumber \\
&-\sum_{\sigma\sigma '}\left[{f}^{\sigma}_{\beta}(1-{f}^{\sigma '}_{\beta})+{f}^{\sigma'}_{\beta}(1-{f}^{\sigma}_{\beta})\right]\left[\text{Tr}\left[|\mathbf{S}^{x}_{\alpha\beta}|^{2}-|\mathbf{S}^{y}_{\alpha\beta}|^{2}\pm|\mathbf{S}^{0}_{\alpha\beta}|^{2}\mp|\mathbf{S}^{z}_{\alpha\beta}|^{2}\right]\right]\nonumber \\
&+\sum_{\gamma\delta\sigma\sigma '}\left[{f}^{\sigma}_{\gamma}(1-{f}^{\sigma '}_{\delta})+{f}^{\sigma '}_{\delta}(1-{f}^{\sigma}_{\gamma})\right]\Big[\text{Tr}\Big[\mathbf{S}_{\alpha\gamma\delta, \beta\delta\gamma}\pm\mathbf{T}_{\alpha\gamma\delta, \beta\delta\gamma}\Big]\Big]\Bigg]
\end{align}

We expressed the charge current noise and spin current noise along transverse and longitudinal direction. We observed that the equilibrium like contribution does not consist of the spin-space entanglement terms. The interference effects as well as spin-space entanglements is observed in the transport like fluctuation terms in both the self and cross correlations, where the former effect wil be observed in next section. It can be observed that the equilibrium and non-equilibrium parts cannot be separated as well as the noise spectra cannot be written simply as a sum of Fermi distribution functions. This is because spin current is non-linear in nature\cite{pareek}. As we can observe from the expressions for noise spectra, every possible interfering term is involved in all the expressions for noise spectra. It can be reasoned that certain combinations of same entangled as well as unentangled terms determine the direction and nature of noise spectra. A specific combination of these terms correspond to shot noise along particular direction. It is the entanglement effect that determines the nature of shot noise in a transverse or longitudinal direction. Even if we consider a unitary scattering matrix, the entanglement between spin and orbital degrees of freedom would be observed but it wouldnot be as pronounced as in the non-unitary case.

The current cross-correlations obtained for zero frequency condition can be utilised to demonstrate the violation of Bell's inequality. We can represent current cross-correlation obtained in equation [12] for a multichannel case. We will be requiring some rational function of the current cross-correlations in order to obtain Bell-CHSH parameters. We take;

\begin{equation}
E=\Bigg[\frac{\mathbf{\mathcal{S}}^{\sigma\sigma , q}_{\alpha\beta}+\mathbf{\mathcal{S}}^{\sigma '\sigma ', q}_{\alpha\beta}-\mathbf{\mathcal{S}}^{\sigma\sigma ', q}_{\alpha\beta}-\mathbf{\mathcal{S}}^{\sigma '\sigma , q}_{\alpha\beta}}{\mathbf{\mathcal{S}}^{\sigma\sigma , q}_{\alpha\beta}+\mathbf{\mathcal{S}}^{\sigma '\sigma ', q}_{\alpha\beta}+\mathbf{\mathcal{S}}^{\sigma\sigma ', q}_{\alpha\beta}+\mathbf{\mathcal{S}}^{\sigma '\sigma , q}_{\alpha\beta}}\Bigg]
\end{equation}
Here $\mathbf{\mathcal{S}}^{\sigma\sigma , q}_{\alpha\beta}$ gives the cross-correlations between currents at the leads $\alpha$ and $\beta$ with spin $\sigma$ or $\sigma '$. The current with respective spin is detected at the normal leads. 
For the detection of Bell's inequality one requires to locally mix the scattering matrices by undergoing unitary transformation. Now, we have the Bell-CHSH parameter given by;
\begin{equation}
\xi=E(\phi_{\alpha},\phi_{\beta})+E(\phi_{\alpha}^{'},\phi_{\beta})+E(\phi_{\alpha},\phi_{\beta}^{'})-E(\phi_{\alpha}^{'},\phi_{\beta}^{'})
\end{equation}
The Bell's inequality gets violated if for some set of unitary transformation, we get $\xi>2$. So, we suggest a method for the generation and detection of spin-orbit entangled electrons in a spin-resolved mesoscopic conductors\cite{loss, Samuelsson, Beenakker}. Reference \cite{Samuelsson} has previously utilised this method to study the entanglement of orbital state in an spin-independent system. In our case, we take exact spin-resolved scattering matrices to generate spin-space entangled electrons and propose their detection in terms of zero frequency current cross-correlators. The entanglement between spin and orbital state survives even for a system determined by unitary scattering matrices but we will not observe a full-fledged entanglement of all possible degrees of freedom in that case. The application of exact scattering matrix ensures all possible entanglement between different spin and orbital states.
\ \\

Further, the decoherence in scattered system can be observed in terms of spin-space entanglement effects\cite{branislav}. It can be observed that, though in a single channel transportation the Dyakonov Perel type decoherence effect vanishes, the entanglement effect still remains\cite{zarbo}. The above formulation can be well extended incorporating the spin density matrix to express the noise and current in a spin dependent system, when a polarised charge current is injected.

\section*{4. Charge and Spin Current Shot Noise in Terms of Spin Flip and Non-Spin Flip Transition Amplitudes}
In this section, we will represent the scattering matrix in terms of spin flip and non-spin flip scattering amplitudes. We have observed quantum interference effect in the expression for spin current\cite{pareek2}, so it can be expected that the interference between the spin flip and non-spin flip scattering amplitudes should produce a significant effect upon the noise spectra. In this section we will try to demonstrate the nature and extent of this effect. Following the same procedure of previous section, under equilibrium condition ($\alpha=\beta$), the self correlation between charge current can be expressed as;
\begin{align}
\boldsymbol{\mathcal{S}}^{\sigma\sigma ', q}_{\alpha\alpha}
&=\frac{{e}^{2}}{{h}}\int dE\Bigg[ \sum_{\sigma\sigma '}\Big[\Big[\mathbf{N}_{\alpha}-2{R}^{\sigma\sigma'}_{\alpha\alpha}\Big]\Big[{f}^{\sigma}_{\alpha}(1-{f}^{\sigma '}_{\alpha})+{f}^{\sigma '}_{\alpha}(1-{f}^{\sigma }_{\alpha})\Big]\Big]+ \sum_{\gamma\delta\sigma '}\Bigg[\Big[{f}^{\sigma}_{\gamma}(1-{f}^{\sigma '}_{\delta})\nonumber \\
&+{f}^{\sigma '}_{\delta}(1-{f}^{\sigma}_{\gamma})+{f}^{-\sigma}_{\gamma}(1-{f}^{\sigma '}_{\delta})+{f}^{\sigma '}_{\delta}(1-{f}^{-\sigma}_{\gamma})\Big]\bigg[\text{Tr}\bigg[\Big(\mathbf{S}^{{\sigma \sigma '}^{\dagger}}_{\alpha\gamma}\mathbf{S}^{{\sigma \sigma '}}_{\alpha\delta}+\mathbf{S}^{{-\sigma \sigma'}^{\dagger}}_{\alpha\gamma}\mathbf{S}^{{-\sigma \sigma '}}_{\alpha\delta}\Big)\nonumber \\
&\left(\mathbf{S}^{{\sigma \sigma '}^{\dagger}}_{\alpha\delta}\mathbf{S}^{{\sigma \sigma '}}_{\alpha\gamma}+\mathbf{S}^{{-\sigma \sigma '}^{\dagger}}_{\alpha\delta}\mathbf{S}^{{-\sigma \sigma '}}_{\alpha\gamma}\right) +\left(\mathbf{S}^{{\sigma \sigma '}^{\dagger}}_{\alpha\gamma}\mathbf{S}^{{\sigma -\sigma '}}_{\alpha\delta}+\mathbf{S}^{{-\sigma \sigma'}^{\dagger}}_{\alpha\gamma}\mathbf{S}^{{-\sigma -\sigma '}}_{\alpha\delta}\right)\left(\mathbf{S}^{{\sigma -\sigma '}^{\dagger}}_{\alpha\delta}\mathbf{S}^{{\sigma \sigma '}}_{\alpha\gamma}+\mathbf{S}^{{-\sigma -\sigma'}^{\dagger}}_{\alpha\delta}\mathbf{S}^{{-\sigma \sigma '}}_{\alpha\gamma}\right)\bigg]\bigg]\Bigg]\Bigg]
\end{align}
As we know the scattering matrix is non-unitary implying non-abelian spin-space entangled system. As a consequence the self correlation in a spin resolved system considerably differs from the spin independent one. The terms giving the transport fluctuations are also present in above expression for equilibrium fluctuations. Hence the separation of equilibrium and transport contributions cannot be attained in spin resolved system as was done in spin independent one\cite{buttiker1}. This difference in behavior of spin resolved system can be attributed to non-vanishing contribution from spin-space entanglement and quantum interference between spin dependent and spin independent transition amplitude, i.e. to the non-unitarity of scattering matrix. It has been observed that the study of noise spectra in spin resolved system requires both the self and cross correlations\cite{wang}. The relation for cross-correlation has been obtained in Appendix B. Moreover, the spin current is of vectorial nature and encompass three out of four components of the complex quaternion representing the spin and charge current in spin resolved system. The different correlations can be divided into transverse and longitudinal components. The longitudinal component of spin current noise under equilibrium condition is obtained as;
\begin{align}
\boldsymbol{\mathcal{S}}^{\sigma \sigma ', z}_{\alpha \alpha}
&= \frac{h}{16{\pi}^{2}} \int dE \Biggl[\left[\sum_{\sigma \sigma '}\left[{f}^{\sigma}_{\alpha}(1-{f}^{\sigma '}_{\alpha})+{f}^{\sigma '}_{\alpha}(1-{f}^{\sigma }_{\alpha})\right]\right]\Big[\mathbf{N}_{\alpha}-2{R}^{\sigma \sigma}_{\alpha \alpha}+2{R}^{-\sigma \sigma}_{\alpha \alpha}+2{R}^{\sigma -\sigma}_{\alpha \alpha}-2{R}^{-\sigma -\sigma}_{\alpha \alpha}\Big]\nonumber \\
&+\sum_{\gamma\delta\sigma '}\Bigg[\left[{f}^{\sigma}_{\gamma}(1-{f}^{\sigma '}_{\delta})+{f}^{\sigma '}_{\delta}(1-{f}^{\sigma}_{\gamma})+{f}^{-\sigma}_{\gamma}(1-{f}^{\sigma '}_{\delta})+{f}^{\sigma '}_{\delta}(1-{f}^{-\sigma}_{\gamma})\right]\bigg[\text{Tr}\bigg[\left(\mathbf{S}^{{\sigma \sigma '}^{\dagger}}_{\alpha\gamma}\mathbf{S}^{{\sigma \sigma '}}_{\alpha\delta}-\mathbf{S}^{{-\sigma \sigma'}^{\dagger}}_{\alpha\gamma}\mathbf{S}^{{-\sigma \sigma '}}_{\alpha\delta}\right) \nonumber \\
&\left(\mathbf{S}^{{\sigma \sigma '}^{\dagger}}_{\alpha\delta}\mathbf{S}^{{\sigma \sigma '}}_{\alpha\gamma}-\mathbf{S}^{{-\sigma \sigma'}^{\dagger}}_{\alpha\delta}\mathbf{S}^{{-\sigma \sigma '}}_{\alpha\gamma}\right) +\left(\mathbf{S}^{{\sigma \sigma '}^{\dagger}}_{\alpha\gamma}\mathbf{S}^{{\sigma -\sigma '}}_{\alpha\delta}-\mathbf{S}^{{-\sigma \sigma'}^{\dagger}}_{\alpha\gamma}\mathbf{S}^{{-\sigma -\sigma '}}_{\alpha\delta}\right)\left(\mathbf{S}^{{\sigma -\sigma '}^{\dagger}}_{\alpha\delta}\mathbf{S}^{{\sigma \sigma '}}_{\alpha\gamma}-\mathbf{S}^{{-\sigma -\sigma'}^{\dagger}}_{\alpha\delta}\mathbf{S}^{{-\sigma \sigma '}}_{\alpha\gamma}\right)\bigg]\bigg]\Bigg]\Biggr]
\end{align}
We have written the correlation between spin currents along longitudinal direction in terms of transmission and reflection probabilities, though the interfering terms are complicated enough to be represented in terms of transition probabilities; we have expressed them in terms of block scattering matrix. We observe that the spin noise spectra consists of equilibrium and non-equilibrium contributions which cannot be separated and also involves the quantum interference between spin dependent and spin independent transition amplitudes. Hence, the total spin noise doesn't vanish even under equilibrium condition. Under time reversal symmetry, $-{R}^{\sigma\sigma}_{\alpha\alpha}+{R}^
{-\sigma\sigma}_{\alpha\alpha}+{R}^
{\sigma-\sigma}_{\alpha\alpha}-{R}^
{-\sigma-\sigma}_{\alpha\alpha}=-2{R}^{\sigma\sigma}
_{\alpha\alpha}+{R}^
{-\sigma\sigma}_{\alpha\alpha}+{R}^
{\sigma-\sigma}_{\alpha\alpha}$ and $\mathcal{T}^{s}_{\alpha\beta}=\mathcal{T}^{s}_{\beta\alpha}$
 where, $\mathcal{T}^{s}_{\alpha\beta}=-{T}^
 {\sigma\sigma}_{\alpha\beta}+{T}^
{-\sigma\sigma}_{\alpha\beta}+{T}^
{\sigma-\sigma}_{\alpha\beta}-{T}^
{-\sigma-\sigma}_{\alpha\beta}$. Hence the coefficient of transmission probability may be written in terms of sum of distribution functions for $\alpha$ and $\beta$ lead. We observe that total reflection probability for time reversal symmetric system is considerably different from that of time reversal asymmetric system.\ \\

It has been shown that transverse shot noise can be utilised to separate different type of SO interactions driving the SHE\cite{dragomirova}. The expression for transverse shot noise is deduced in Appendix B. The magnetisation noise from shot noise can be used to classify the role of mixing conductance in terms of interference and entanglement effects. The fluctuation in magnetisation vector, thus is dependent on the entanglement effects and can be clarified through above formulation. Hence we observe that both the transverse and longitudinal component of shot noise have their respective significance.

\section*{5. Summary} 
In summary, we observed the entanglement effect between spin and orbital degrees of freedom in a spin-resolved mesoscopic system. The entanglement was observed in the expression for current as well as shot noise. Hence, our main result is the generation of spin-orbit entangled electrons in a spin-resolved mesoscopic conductor. We have calculated the zero frequency current correlations which has been utilised for the detection of entanglement. Besides that, we have studied the spin-resolved scattering in a mesoscopic conductor in its generality. Moreover, the quantum interference was observed in between spin flip and non-spin flip scattering amplitudes. The proposed expression for shot noise of spin and charge current exhibit quantum entanglement as well as quantum interference effects. Further, it has been observed that shot noise for transverse and longitudinal direction depend upon different combination of entangled and unentangled terms. So, the entanglement effect plays a primary role in determining the nature of current and noise in a spin-resolved mesoscopic system. In some cases, where quantum interference between spin flip and non-spin flip transition amplitude occurs predominantly, the noise spectrum cannot be written in terms of transmission and reflection probabilities, so we have used the block scattering matrix for the representation. The quantum interference between spin flip and non-spin flip amplitudes also give mixed conductance as observed in reference \cite{brataas}.\\
\ \\

 \section*{Appendix}
 \subsection*{Appendix A}
  
 To obtain the expression for charge current in equation [3], we did some calculations.
  \begin{align}
  \sum_{\beta\gamma}\langle\boldsymbol{A}^{\dagger}_{\beta}(E)\boldsymbol{M}_{\beta\gamma}(\alpha, E)\boldsymbol{A}_{\gamma}(E)\rangle
  &= \sum_{\beta}\text{Tr}\left[\boldsymbol{M}_{\beta\beta}(\alpha, E)\right]{f}_{\beta}(E)\nonumber \\
  &= \sum_{\beta}\text{Tr}\left[\Lambda_{\alpha}\delta_{\alpha\beta}-
  \boldsymbol{S}^{\dagger}_{\alpha\beta}\Lambda_{\alpha}
  \boldsymbol{S}_{\alpha\beta}\right]{f}_{\beta}(E)
 \end{align} 
 where,
 \[\mathbf{M}_{\beta\gamma}(\alpha,E)=\mathbf{\Lambda}_{\alpha}\delta_{\alpha\beta}\delta_{\alpha\gamma}-\mathbf{S}^{\dagger}_{\alpha\beta}(E)\mathbf{\Lambda}_{\alpha}\mathbf{S}_{\alpha\gamma}(E)\]

 Similarly, we replace $\boldsymbol{M}$ with $\boldsymbol{\Gamma}$ and proceed as above to obtain average spin current.

 where, \[\boldsymbol{\Gamma}^{i}_{\beta\gamma}=\mathbf{\Omega}_{\alpha,i}\delta_{\alpha\beta}\delta_{\alpha\gamma}-\mathbf{S}^{\dagger}_{\alpha\beta}(E)\mathbf{\Omega}_{\alpha, i}\mathbf{S}_{\alpha\gamma}(E)\]
\[\mathbf{\Omega}_{\alpha}=\boldsymbol{\sigma}\otimes\mathbf{I}_{\alpha}\]

  We obtain,
  \begin{align}
  \sum_{\beta\gamma}\langle\boldsymbol{A}^{\dagger}_{\beta}(E)\boldsymbol{\Gamma}^{i}_{\beta\gamma}(\alpha, E)\boldsymbol{A}_{\gamma}(E)\rangle
   &= \sum_{\beta}\text{Tr}\left[\Omega^{i}_{\alpha}\delta_{\alpha\beta}-
  \boldsymbol{S}^{\dagger}_{\alpha\beta}\Omega^{i}_{\alpha}
  \boldsymbol{S}_{\alpha\beta}\right]{f}_{\beta}(E)
 \end{align} 
 
 Moreover, we have used the following calculations in expressing the shot noise;
 \begin{align}
 \sum_{\gamma\delta}\text{Tr}\left[\boldsymbol{\Gamma}^{i}_{
 \gamma\delta}(\alpha, E)\boldsymbol{\Gamma}^{i}_{\delta\gamma}(\beta, E)\right]
 &=\sum_{\gamma\delta}\text{Tr}\bigg[\Omega_{\alpha,i}\delta_{\alpha\gamma}\delta_
 {\alpha\delta}\Omega_{\beta,i}\delta_{\beta\delta}\delta_
 {\beta\gamma}- \Omega_{\alpha,i}\delta_{\alpha\gamma}\delta_
 {\alpha\delta}\boldsymbol{S}^{\dagger}_{\beta\delta}
 \Omega_{\beta,i}\boldsymbol{S}_{\beta\gamma}\nonumber \\
 &-
 \boldsymbol{S}^{\dagger}_{\alpha\gamma}
 \Omega_{\alpha,i}\boldsymbol{S}_{\alpha\delta}
 \Omega_{\beta,i}\delta_{\beta\delta}\delta_
 {\beta\gamma}+ \boldsymbol{S}^{\dagger}_{\alpha\gamma}
 \Omega_{\alpha,i}\boldsymbol{S}_{\alpha\delta}\boldsymbol{S}^{\dagger}_{\beta\delta}
 \Omega_{\beta,i}\boldsymbol{S}_{\beta\gamma}\bigg]
\end{align}  
The first term in the right hand side of equation [32] survives only when $\alpha=\beta$, i.e. it contributes only to self-correlation.\\
\ \\
\subsection*{Appendix B}
 The equilibrium-like cross-correlation between charge current can be obtained by taking $\alpha\neq\beta$; we consider different chemical potential for different contacts. 
\begin{align}
\boldsymbol{\mathcal{S}}^{\sigma\sigma ', q}_{\alpha\beta, \alpha\neq\beta}
&=\frac{{e}^{2}}{{h}}\int dE \Bigg[\sum_{\sigma\sigma '}\Big[-{T}^{\sigma\sigma '}_{\beta\alpha}\Big[{f}^{\sigma}_{\alpha}(1-{f}^{\sigma '}_{\alpha})+{f}^{\sigma '}_{\alpha}(1-{f}^{\sigma }_{\alpha})\Big]-{T}^{\sigma\sigma '}_{\alpha\beta}\Big[{f}^{\sigma}_{\beta}(1-{f}^{\sigma '}_{\beta})+{f}^{\sigma '}_{\beta}(1-{f}^{\sigma }_{\beta})\Big]\Big]\nonumber \\
&+\sum_{\gamma\delta\sigma '}\Bigg[\left[{f}^{\sigma}_{\gamma}(1-{f}^{\sigma '}_{\delta})+{f}^{\sigma '}_{\delta}(1-{f}^{\sigma}_{\gamma})+{f}^{-\sigma}_{\gamma}(1-{f}^{\sigma '}_{\delta})+{f}^{\sigma '}_{\delta}(1-{f}^{-\sigma}_{\gamma})\right]\bigg[\text{Tr}\bigg[\Big(\mathbf{S}^{{\sigma \sigma '}^{\dagger}}_{\alpha\gamma}\mathbf{S}^{{\sigma \sigma '}}_{\alpha\delta}+\mathbf{S}^{{-\sigma \sigma'}^{\dagger}}_{\alpha\gamma}\mathbf{S}^{{-\sigma \sigma '}}_{\alpha\delta}\Big) \nonumber \\
&\left(\mathbf{S}^{{\sigma \sigma '}^{\dagger}}_{\beta\delta}\mathbf{S}^{{\sigma \sigma '}}_{\beta\gamma}+\mathbf{S}^{{-\sigma \sigma '}^{\dagger}}_{\beta\delta}\mathbf{S}^{{-\sigma \sigma '}}_{\beta\gamma}\right) +\left(\mathbf{S}^{{\sigma \sigma '}^{\dagger}}_{\alpha\gamma}\mathbf{S}^{{\sigma -\sigma '}}_{\alpha\delta}+\mathbf{S}^{{-\sigma \sigma'}^{\dagger}}_{\alpha\gamma}\mathbf{S}^{{-\sigma -\sigma '}}_{\alpha\delta}\right)
\left(\mathbf{S}^{{\sigma -\sigma '}^{\dagger}}_{\beta\delta}\mathbf{S}^{{\sigma \sigma '}}_{\beta\gamma}+\mathbf{S}^{{-\sigma -\sigma'}^{\dagger}}_{\beta\delta}\mathbf{S}^{{-\sigma \sigma '}}_{\beta\gamma}\right)\bigg]\bigg]\Bigg]\Bigg]
\end{align}
The expression for noise power spectrum of charge current in equation [24] have both interfering and non-interfering terms. The interference is observed in terms of mixing of spin flip and non-spin flip scattering amplitudes. For time reversal symmetric system, we follow the same line of argument, done for the case of charge and spin current.
Now, the cross-correlation between spin current along longitudinal direction can be expressed as;
\begin{align}
\boldsymbol{\mathcal{S}}^{\sigma \sigma ', z}_{\alpha \beta, \alpha\neq\beta}
&= \frac{h}{16{\pi}^{2}} \int dE \Biggl[\left[\sum_{\sigma \sigma '}\left[{f}^{\sigma}_{\alpha}(1-{f}^{\sigma '}_{\alpha})+{f}^{\sigma '}_{\alpha}(1-{f}^{\sigma }_{\alpha})\right]\right]\Big[-{T}^{\sigma \sigma}_{\beta \alpha}
+{T}^{-\sigma \sigma}_{\beta \alpha}+{T}^{\sigma -\sigma}_{\beta \alpha}-{T}^{-\sigma -\sigma}_{\beta \alpha}\Big]\nonumber \\
&+\left[\sum_{\sigma \sigma '}\left[{f}^{\sigma}_{\beta}(1-{f}^{\sigma '}_{\beta})+{f}^{\sigma '}_{\beta}(1-{f}^{\sigma }_{\beta})\right]\right]\left[-{T}^{\sigma \sigma}_{\alpha \beta}+{T}^{-\sigma \sigma}_{\alpha \beta}+{T}^{\sigma -\sigma}_{\alpha \beta}-{T}^{-\sigma -\sigma}_{\alpha \beta}\right]+\sum_{\gamma\delta\sigma '}\Bigg[\Big[{f}^{\sigma}_{\gamma}(1-{f}^{\sigma '}_{\delta})\nonumber \\
&+{f}^{\sigma '}_{\delta}(1-{f}^{\sigma}_{\gamma})+{f}^{-\sigma}_{\gamma}(1-{f}^{\sigma '}_{\delta})+{f}^{\sigma '}_{\delta}(1-{f}^{-\sigma}_{\gamma})\Big]\bigg[\text{Tr}\bigg[\left(\mathbf{S}^{{\sigma \sigma '}^{\dagger}}_{\alpha\gamma}\mathbf{S}^{{\sigma \sigma '}}_{\alpha\delta}-\mathbf{S}^{{-\sigma \sigma'}^{\dagger}}_{\alpha\gamma}\mathbf{S}^{{-\sigma \sigma '}}_{\alpha\delta}\right) \nonumber \\
&\left(\mathbf{S}^{{\sigma \sigma '}^{\dagger}}_{\beta\delta}\mathbf{S}^{{\sigma \sigma '}}_{\beta\gamma}-\mathbf{S}^{{-\sigma \sigma '}^{\dagger}}_{\beta\delta}\mathbf{S}^{{-\sigma \sigma '}}_{\beta\gamma}\right) +\left(\mathbf{S}^{{\sigma \sigma '}^{\dagger}}_{\alpha\gamma}\mathbf{S}^{{\sigma -\sigma '}}_{\alpha\delta}-\mathbf{S}^{{-\sigma \sigma'}^{\dagger}}_{\alpha\gamma}\mathbf{S}^{{-\sigma -\sigma '}}_{\alpha\delta}\right)\left(\mathbf{S}^{{\sigma -\sigma '}^{\dagger}}_{\beta\delta}\mathbf{S}^{{\sigma \sigma '}}_{\beta\gamma}-\mathbf{S}^{{-\sigma -\sigma'}^{\dagger}}_{\beta\delta}\mathbf{S}^{{-\sigma \sigma '}}_{\beta\gamma}\right)\bigg]\bigg]\Bigg]
\end{align}

 Now, the spin shot noise in X and Y directions can be obtained as;
\begin{align}
\boldsymbol{\mathcal{S}}^{\sigma \sigma ', x, y}_{\alpha \alpha}
&= \frac{h}{16{\pi}^{2}} \int dE \Bigg[\left[\sum_{\sigma \sigma '}\Big[{f}^{\sigma}_{\alpha}(1-{f}^{\sigma '}_{\alpha})+{f}^{\sigma '}_{\alpha}(1-{f}^{\sigma }_{\alpha})\Big]\right]\bigg[\mathbf{N}_{\alpha}\mp\bigg[\text{Tr}\bigg[2{\mathbf{S}}^{{-\sigma \sigma}^{\dagger}}_{\alpha\alpha}{\mathbf{S}}^{\sigma-\sigma}_{\alpha\alpha}\pm2{\mathbf{S}}^{{\sigma \sigma}^{\dagger}}_{\alpha\alpha}{\mathbf{S}}^{-\sigma-\sigma}_{\alpha\alpha}\nonumber \\
&\pm2{\mathbf{S}}^{{-\sigma -\sigma}^{\dagger}}_{\alpha\alpha}{\mathbf{S}}^{\sigma\sigma}_{\alpha\alpha}+2{\mathbf{S}}^{{\sigma -\sigma}^{\dagger}}_{\alpha\alpha}{\mathbf{S}}^{-\sigma\sigma}_{\alpha\alpha}\bigg]\bigg]\bigg]\pm\sum_{\gamma\delta\sigma '}\Bigg[\left[{f}^{\sigma}_{\gamma}(1-{f}^{\sigma '}_{\delta})+{f}^{\sigma '}_{\delta}(1-{f}^{\sigma}_{\gamma})+{f}^{-\sigma}_{\gamma}(1-{f}^{\sigma '}_{\delta})+{f}^{\sigma '}_{\delta}(1-{f}^{-\sigma}_{\gamma})\right]\nonumber \\
 &\bigg[\text{Tr}\bigg[\left(\mathbf{S}^{{-\sigma \sigma '}^{\dagger}}_{\alpha\gamma}\mathbf{S}^{{\sigma \sigma '}}_{\alpha\delta}\pm\mathbf{S}^{{\sigma \sigma'}^{\dagger}}_{\alpha\gamma}\mathbf{S}^{{-\sigma \sigma '}}_{\alpha\delta}\right)\left(\mathbf{S}^{{-\sigma \sigma '}^{\dagger}}_{\alpha\delta}\mathbf{S}^{{\sigma \sigma '}}_{\alpha\gamma}\pm\mathbf{S}^{{\sigma \sigma '}^{\dagger}}_{\alpha\delta}\mathbf{S}^{{-\sigma \sigma '}}_{\alpha\gamma}\right)+\left(\mathbf{S}^{{-\sigma \sigma '}^{\dagger}}_{\alpha\gamma}\mathbf{S}^{{\sigma -\sigma '}}_{\alpha\delta}\pm\mathbf{S}^{{\sigma \sigma'}^{\dagger}}_{\alpha\gamma}\mathbf{S}^{{-\sigma -\sigma '}}_{\alpha\delta}\right)\nonumber \\
 &\left(\mathbf{S}^{{-\sigma -\sigma '}^{\dagger}}_{\alpha\delta}\mathbf{S}^{{\sigma \sigma '}}_{\alpha\gamma}\pm\mathbf{S}^{{\sigma -\sigma'}^{\dagger}}_{\alpha\delta}\mathbf{S}^{{-\sigma \sigma '}}_{\alpha\gamma}\right)\bigg]\bigg]\Bigg]\Bigg]
\end{align}
Further, the cross-correlation along the transverse direction can be obtained as;
\begin{align}
\boldsymbol{\mathcal{S}}^{\sigma \sigma ', x, y}_{\alpha \beta}
&= \frac{h}{16{\pi}^{2}} \int dE \Bigg[\left[\sum_{\sigma \sigma '}\Big[{f}^{\sigma}_{\alpha}(1-{f}^{\sigma '}_{\alpha})+{f}^{\sigma '}_{\alpha}(1-{f}^{\sigma }_{\alpha})\Big]\right]\bigg[\mp\text{Tr}\bigg[{\mathbf{S}}^{{\sigma -\sigma}^{\dagger}}_{\beta\alpha}{\mathbf{S}}^{-\sigma\sigma}_{\beta\alpha}\pm{\mathbf{S}}^{{-\sigma -\sigma}^{\dagger}}_{\beta\alpha}{\mathbf{S}}^{\sigma\sigma}_{\beta\alpha}\pm{\mathbf{S}}^{{\sigma \sigma}^{\dagger}}_{\beta\alpha}{\mathbf{S}}^{-\sigma-\sigma}_{\beta\alpha}\nonumber \\
&+{\mathbf{S}}^{{-\sigma \sigma}^{\dagger}}_{\beta\alpha}{\mathbf{S}}^{\sigma-\sigma}_{\beta\alpha}\bigg]\bigg]\mp\left[\sum_{\sigma \sigma '}\Big[{f}^{\sigma}_{\beta}(1-{f}^{\sigma '}_{\beta})+{f}^{\sigma '}_{\beta}(1-{f}^{\sigma }_{\beta})\Big]\right]\bigg[\text{Tr}\bigg[{\mathbf{S}}^{{-\sigma \sigma}^{\dagger}}_{\alpha\beta}{\mathbf{S}}^{\sigma-\sigma}_{\alpha\beta}\pm{\mathbf{S}}^{{\sigma \sigma}^{\dagger}}_{\alpha\beta}{\mathbf{S}}^{-\sigma-\sigma}_{\alpha\beta}\pm{\mathbf{S}}^{{-\sigma -\sigma}^{\dagger}}_{\alpha\beta}{\mathbf{S}}^{\sigma\sigma}_{\alpha\beta}\nonumber \\
&+{\mathbf{S}}^{{\sigma -\sigma}^{\dagger}}_{\alpha\beta}{\mathbf{S}}^{-\sigma\sigma}_{\alpha\beta}\bigg]\bigg]\pm\sum_{\gamma\delta\sigma '}\Bigg[\left[{f}^{\sigma}_{\gamma}(1-{f}^{\sigma '}_{\delta})+{f}^{\sigma '}_{\delta}(1-{f}^{\sigma}_{\gamma})+{f}^{-\sigma}_{\gamma}(1-{f}^{\sigma '}_{\delta})+{f}^{\sigma '}_{\delta}(1-{f}^{-\sigma}_{\gamma})\right]\nonumber \\
 &\bigg[\text{Tr}\bigg[\left(\mathbf{S}^{{-\sigma \sigma '}^{\dagger}}_{\alpha\gamma}\mathbf{S}^{{\sigma \sigma '}}_{\alpha\delta}\pm\mathbf{S}^{{\sigma \sigma'}^{\dagger}}_{\alpha\gamma}\mathbf{S}^{{-\sigma \sigma '}}_{\alpha\delta}\right)\left(\mathbf{S}^{{-\sigma \sigma '}^{\dagger}}_{\beta\delta}\mathbf{S}^{{\sigma \sigma '}}_{\beta\gamma}\pm\mathbf{S}^{{\sigma \sigma '}^{\dagger}}_{\beta\delta}\mathbf{S}^{{-\sigma \sigma '}}_{\beta\gamma}\right)+\left(\mathbf{S}^{{-\sigma \sigma '}^{\dagger}}_{\alpha\gamma}\mathbf{S}^{{\sigma -\sigma '}}_{\alpha\delta}\pm\mathbf{S}^{{\sigma \sigma'}^{\dagger}}_{\alpha\gamma}\mathbf{S}^{{-\sigma -\sigma '}}_{\alpha\delta}\right)\nonumber \\
 &\left(\mathbf{S}^{{-\sigma -\sigma '}^{\dagger}}_{\beta\delta}\mathbf{S}^{{\sigma \sigma '}}_{\beta\gamma}\pm\mathbf{S}^{{\sigma -\sigma'}^{\dagger}}_{\beta\delta}\mathbf{S}^{{-\sigma \sigma '}}_{\beta\gamma}\right)\bigg]\bigg]\Bigg]\Bigg]
\end{align}
 
The upper and lower sign in equations [28] and [29] are for -xx and -yy components respectively, and any single sign are for both components.

\end{document}